# Opportunities and Challenges for 3D Systems and Their Design

Philip Emma and Eren Kursun
IBM Research

> *Editor's note:*
> This article presents the system design opportunities offered by 3D integration, and it discusses the design and test challenges for 3D ICs, with various new design-for-manufacture and DFT issues.
> —Yuan Xie, Pennsylvania State University

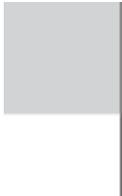

ALTHOUGH IT'S NOT A NEW CONCEPT, 3D integration increasingly receives widespread interest and focus as lithographic scaling becomes more challenging, and as the ability to make miniature vias greatly improves. Like Moore's law, 3D integration improves density. With improvements in packaging density, however, come the challenges associated with its inherently higher power density. And though it acts somewhat as a scaling accelerator, the vertical inte- gration also poses new challenges to design and manufacturing technologies.

The placement of circuits, vias, and macros in the planes of a 3D stack must be co-designed across layers (or must conform to new standards) so that, when assembled, they have correct spatial correspondence. Each layer, although perhaps being a mere functional slice through a system (and we can slice the system in many different ways), must be independently testable so that we can systematically test and diagnose subsystems before and after final assembly. When those layers are assembled, they must come together in a way that enables a sensible yield and facilitates testing the finished product.

To make the most of 3D integration, we should articulate the leverages of 3D systems (other researchers offer a more complete treatment elsewhere[1-5]). Then we can enumerate and elucidate many of the new challenges posed by the design, assembly, and test of 3D systems.

## Unique Leverages of 3D Integration

Although 3D integration affords the same gross benefits as Moore's law in circuit density, it's worth mentioning two imminent concerns about such scaling. We take no sides on either issue; we merely point out that where one stands on these concerns will color one's perception on the advantages of 3D integration.

First, as devices become smaller, three things happen: the device performance doesn't scale, device leakage increases, and device variability worsens. Using 3D integration, however, allows better density without making the devices smaller.

Second, the lithography needed to make devices significantly smaller is costly and even questionable. Yet 3D integration does not require smaller devices; it works independently of, and in synergy with, this kind of scaling.[1,6] Besides raw density, 3D integration allows five new degrees of freedom; none of them are useful in all market sectors, but each is useful in some. Whether they represent a real opportunity to you depends on exactly what you're trying to do, and the reason that you're trying to do it. Those considerations strongly influence how you should choose to practice the art of 3D design. First, by integrating multiple components into a single stack (that is, a single component) 3D integration enables a simpler package to suffice, and it simplifies the subsequent assembly processes to make the end product. This could represent a significant cost advantage, assuming that the cost of the 3D component is reasonable, and that the volumes are at a level sufficient for amortizing the nonrecoverable engineering (NRE) costs of 3D. Second, the modular integration of layers can enable a range of products to be made from a common set of subsystems (where we consider each layer a subsystem). This has the effect of volumizing those subsystems (which reduces cost), and it simplifies the overall design effort associated with that range of products (which also reduces cost). This is exactly the philosophy behind ASIC books, but here we practice it directly at the physical level.

A third consideration is that 3D integration allows us to potentially combine disparate technologies within a single stacked component——such as DRAM with high-speed logic——in a manner that doesn't compromise either technology. It could also allow us to combine an older technology node with a newer technology, which could save cost and schedule on new products if it allowed for the direct reuse of system parts that don't (particularly) need updating for the new product. It could also enable a simpler and lower power integration of communications subsystems——such as silicon-germanium (SiGe) technology, gallium arsenide (GaAs) semiconductors, and optoelectronics——as well as accelerators.

A fourth factor is that, with 3D integration, we can incorporate pieces of the electrical and service infrastructures directly for much better electrical performance. For instance, integrating voltage regulators within a stack delivers cleaner power locally, more efficiently, and more controllably. Furthermore, this can potentially allow for power distribution at a higher voltage and lower current, which gives us greener technology. We could integrate passives more elegantly (for example, by placing decoupling capacitors liberally and locally to stiffen the power rails). Also, we could incorporate clocking and test-related logic in a more modular way.

Finally, with a small via pitch, it's possible to build short and wide vertical buses within a 3D stack. This is useful only if there are elements within the stack that we can place co-spatially, and which would benefit from massive bandwidth. It's this last point that raises issues about how to make vias and how to place them. Of these five leverages, only the last one affords a direct system-level performance advantage, and it does so only for systems that would benefit from a higher internal bandwidth to an integrated (within the stack) cache structure. Integrating cache layers within the stack (instead of connecting to them as off-chip entities) eliminates the slower and higher-power off-chip buses, and instead connects to them with higher-bandwidth, lower-latency vertical buses. Figure 1 shows micro-C4 connections between layers in a stack. Not only does this increase the bandwidth to the cache layers dramatically, it also improves the access latency and lowers the transmission power.

Compared to tens of millimeters per wire to connect logic and memory chips in the 2D plane, within a 3D stack we can connect using through-silicon vias (TSVs) that are mere tens of microns long. This is several orders of magnitude less. Wire-limited performance improvement through vertical integration is projected as the square root of the number of layers in a 3D stack. Other studies show that increasing the number of active layers through vertical integration significantly improves the interconnection performance and bandwidth. Researchers propose various alternatives to interconnect the layers in the stack, including micro-C4 techniques[7] and Cu-Cu thermocompression bonding for bulk.[8]

## Design and Test Challenges in 3D Stacks

Because 3D integration opens up a whole new set of challenges, we should ask ourselves several questions——involving via pitch and placement, thermal issues, and scan chain reconfiguration, among other things——to consider before designing and testing 3D stacks.

### Vias: Pitch and Placement

The first set of questions to answer before building anything in 3D revolves around TSVs, which are essential to 3D stacks because they're the means for interconnecting the layers. The answers to those questions depend heavily on why you're using 3D technology: How much current do you need? How many vias do you need? (And though the holes were rather small, they had to count them all. Now they know how many holes it takes to fill the Albert Hall.) How big do the vias need to be? Where should they be placed? For what are the vias (mostly) used? Of what material are the vias made? What are the vias' electrical, mechanical, and thermal characteristics?

For lower-power applications like commodity DRAM and the commercial mobile space, we need not bring lots of power conduits (vias) through the stack. Further, these applications don't require that many signals. For markets like these, the issues surrounding vias are not paramount and therefore do not heavily influence other design choices. In these markets, the number of vias is small, so their sizes, placements, and constructions don't bear heavily on other design considerations.

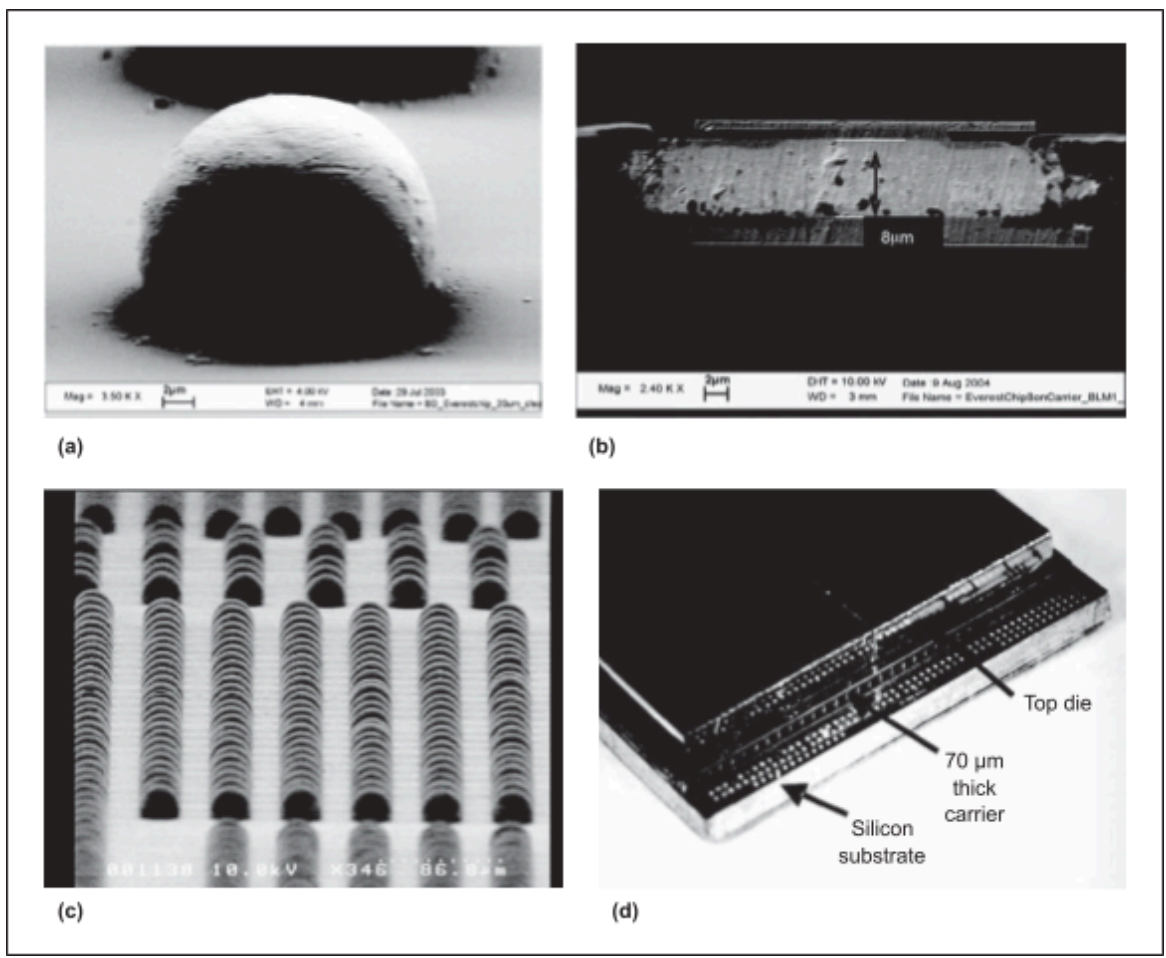

Figure 1. Using micro-C4s provides interconnectivity among layers in the stack. Micro-C4 (a). Cross-section view of a micro-C4 connection (b). A 50-micron pitch micro-C4 array (c). Assembled 3D stack (d).

In high-power applications such as microprocessors, vias are paramount: they drive most of the other design considerations. In applications like this, power needs delivery on a regular grid, and distances between the points on that grid should be short, because we can't distribute high power horizontally on the chip's back-end-of-line (BEOL) wiring. Therefore, we should place power vias regularly on a closed grid. This pitch will decrease with increasing current demand.

Further, power vias must extend through the BEOL wiring layers; they cannot simply use BEOL wiring layers to deliver large currents vertically. Consider a via that's etched through the bulk, and that connects to the next chip layer using the BEOL wiring hierarchy, as Figure 2 shows. While this kind of via is the easiest to make (because we need not deal with extending the via through the BEOL) and suffices for getting signals through a layer, a via structure like this is unsuitable for power delivery.[9]

Therefore, although we might choose to have via structures like the one in Figure 2 for transmitting signals, we also need power vias that cut all the way through the BEOL structure. Although it might make sense to have multiple via types, allowing this can drive more design complexity than defining a single via type (the power via) and using that via for signals, too. When placing power vias in a grid, this tends to constrain the placement of signal vias. In particular, *x* wiring (and *y* wiring) can't run across the chip coincident with the *x* grid (or *y* grid) of power vias. Therefore, it makes the most sense to place groups of signal vias between the power vias, within that grid, as Figure 3 shows. Finally, to connect the chips to the package, and to set an adequate power delivery grid, a number of processes (C4NP).[7] Patel et al.[10] also reported pitches ranging from 225 microns to fewer than 50 microns with TSV heights of 300 microns to fewer than 50 microns. The TSV pitch and size not only determine the kind of 3D integration enabled at the system level, but they impact the stack's physical characteristics. The ability to probe signals becomes more challenging as 3D integration becomes more sophisticated (such as with finer TSV pitches and thinner silicon layers), which will cause more complications in the future.

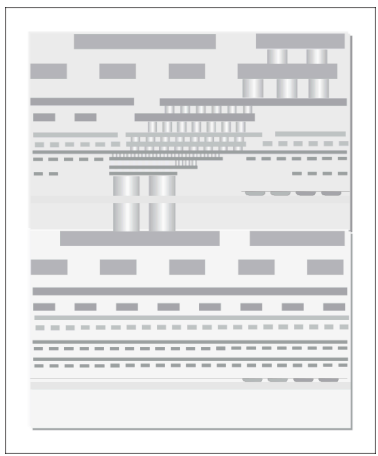

Figure 2. Through-silicon vias (TSVs) that don't extend through the back-end-of-line (BEOL) wiring layers aren't suitable for power delivery.

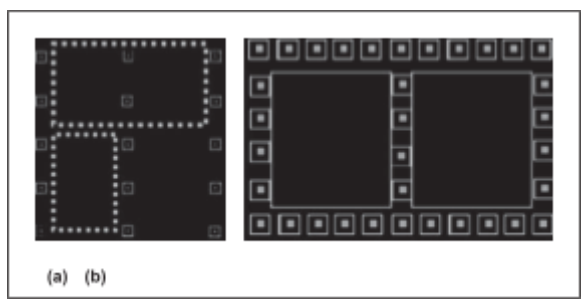

Figure 3. Macro placement constraints for various via pitches. Such macros are problematic to design over a power grid (a). Signal vias should coincide with the power grid, leaving macros intact (b).

## Capturing and Probing Signals

In principle, signals that run between layers can connect arbitrary logic circuits within any latch-tolatch path spread across those layers. For example, the output of a logic gate on one layer might also be the input to a logic gate on another layer. But it's necessary (at this time——within the nascency of 3D systems) to be able to access that signal for testing. To access such a signal for logic testing, we need to either put a latch on both sides of the connecting via, or to connect the via contact to a landing pad that we can access with a test probe. If the signal is a logic-to-logic signal (that is, the signal is not on a latch boundary) and the signal is not amenable to alternating current (AC) testing, then the first solution requires that additional latches and an additional clocking structure be put in place just to capture this midcycle state for logic testing. The latter solution (a test probe) requires that we connect a large capacitance electrostatic discharge (ESD) diode to the signal landing pad to protect the circuits from static charge on the test probe. This will degrade the signal by significantly slowing it. Note that, either way——putting a latch on both sides or connecting the via to a landing pad——having signals cross layers in the middle of a combinational logic flow is problematic when it comes to testing them, so designers should avoid a situation like this. Such approaches require some significant innovations before they become practicable.

A more practicable design should have latches at both sides of any layer-to-layer interface (via). Ideally, both latches are part of the machine's logic flow (for example, in a latch-to-latch path), so that AC effects are not critical to test (assuming that a signal can comfortably traverse layers in a cycle). However, if the path is not strictly latch to latch, but you're sufficiently confident that timing will not be an issue because it will not need to be tested, then you should use boundary-scan latches to capture the layer-to-layer signals during the test. Of course, we can perform rudimentary AC tests with boundary-scan latches on a single layer, but launching a signal from a scan latch on a single layer (in a test) fails to account for the impedance of the via in the final product.

We state axiomatically that signals traversing a 3D structure through vias must be latch-to-latch signals to be testable, although the capturing latches might simply be boundary scan latches that are not part of the nominal machine flow.[11] Having said this, we can now differentiate between those signals that are only accessible through scan rings, and those that a tester can probe.

Most signals in a large system must be of the former type, only because there are too many of them to probe.

Whereas it is necessary to probe some of the signals (or at least, to be able to connect tester probes to an on-chip test infrastructure), any signal being probed requires two things: a landing pad adjacent to the via that will accommodate the test probe; and large-capacitance ESD structure that will protect the circuits from static discharge when we connect the probe. Also note that a test probe having hundreds of contacts will exert quite a bit of force on the chip being tested. If the chip is several hundred microns thick, this is generally not a problem. But if it's only tens of microns thick, the test probe can do real damage. So testing layers prior to thinning them is prudent. However, testing chips in this way will obviously not detect problems created during the thinning process. In either case, test probes will damage the surface of the landing pads by leaving pockmarks and metal shards. We might need to repair (replanarize) the pads prior to stacking the tested chips, or these deformations can cause subsequent problems. A landing pad and an ESD device are big. Having 3D stacks with many signals would be infeasible if all vias required them. Although we need the pads to give the tester accessibility to each layer, clearly most signals passing through vias will be accessible only to scan chains.

## Assembling and Reconfiguring Scan Chains

Although it's possible to improve the yield of a stack by using known-good dies (KGD) in a lab environment, the cost of doing this in production is prohibitive in a high-volume product. In an ideal manufacturing scenario, it's best to stack finished wafers up and then dice them. Practically, we can't hope for all chip sites in all layers in all 3D stacks made in this way to work, which is why it is essential to make all the layers independently testable prior to stacking them. If we do this, then at least we know what to expect when wafers are stacked. When some of the 3D chiplets work incorrectly, we will know why, and whether we can repair them. Therefore, to be cost-effective, it's essential to perform and collect detailed failure data and analysis on each chip on a wafer. It's also essential (in a wafer-to-wafer assembly process) to incorporate enough redundancy into the layers and the system to be able to repair most problems.

As already discussed, each layer should be testable through scan rings. On-layer built-in self-test (BIST) engines will control many of these. All signals that enter and leave a layer (for example, all via points) should be accessible through a boundary scan. When designed in this way, each layer may have many rings, and it may take lots of time to test each layer.

Testing layers in this way might not be the most effective way to test the finished product (the aggregated stack), so we also need a way to test the layer-to-layer paths (the vias) in the finished product. Therefore, when the layers are finally assembled, we might want to reconfigure the scan rings within the aggregated product to enable a final testing procedure that's both more complete and more efficient, which will require considerable forethought.

Partitioning the scan chains effectively through the stack is important because the testable state (number and length of the scan chains) in the finished stack can be large, making the testing more challenging. One approach is to serially connect the scan chains from each of the 2D layers and preserve the existing order for each layer; however, this will lead to long chains that probably aren't optimal. A better approach is to enable scanning across the layer boundaries, and to add some additional infrastructure that lets us access the existing 2D chains hierarchically. Although this approach makes the parts of the finished product more directly accessible for testing, there must be reasonable certainty that the vertical interconnection infrastructure (vias) will work. Clearly, we can increase certainty by using redundant vias and/or redundant scan paths.

The choice of face-to-face (F2F) and face-to-back (F2B) integration is yet another key consideration in manufacturability and testing. Whereas the testing and accessibility of F2F structures has numerous advantages, the F2F scheme cannot be continued beyond two layers without significant complications. With more than two layers, then, usually the face-toback integration step follows face-to-face integration. Note that while (perhaps) mechanically simpler, FTF requires a mirrored design on one layer so that it complies with the other layer, a characteristic that might add complexity to the design tools.

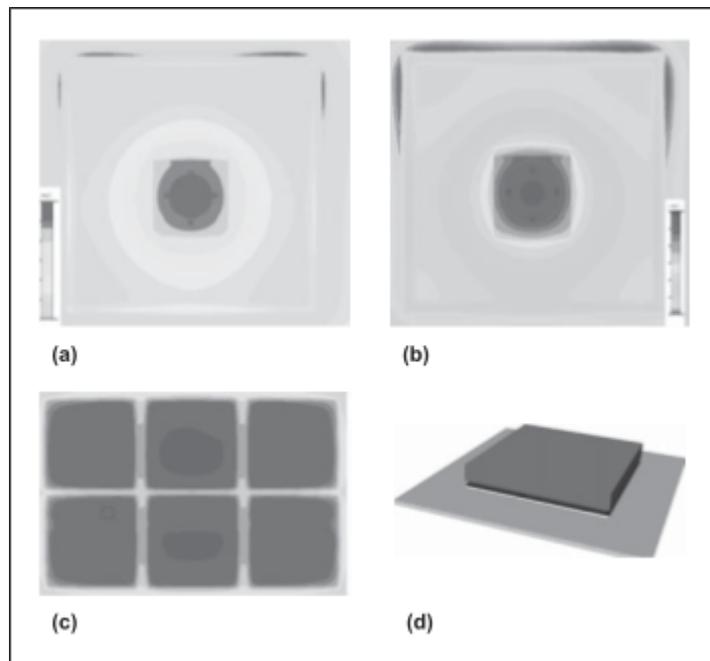

Figure 4. Comparison of 2D (a) and 3D (b) thermal images. Thermal map corresponding to a six-core version for the same power budget (c). Stacked design with the packaging and cooling in place (d).

### Thermal challenges

Thermal challenges limit integration in general and are amplified in 3D technology for high-power applications. Recently, a number of studies showed that the increased thermal density of a two-layer stack is manageable.[12] In addition to the total heat in a stack increasing because of the raw increase in circuit density, spreading those circuits across layers introduces new thermal resistance, which makes the task of removing heat more difficult.

In fact, it can be argued that if the power density in a stack is kept constant as the number of layers increases, the (vertical) thermal resistance will cause ever-increasing thermal gradients. Although a wide range of cooling solutions have been proposed (from standard cold-plate cooling to exotic microchannel schemes), thermal planning and management of the on-chip resources are essential for cost-effective 3D design, no matter what the thermal solution.[13] Ironically, the thinning of layers that is requisite for efficient vertical thermal flow through the stack will impede the horizontal thermal flow within a layer.

The device layers in 3D stacks are generally thinned considerably from the standard layers used in 2D designs. Figures 4a and 4b show a comparison between 2D and 3D thermal images for the total dissipation implemented in a traditional 2D design versus a 3D stack. (We incorporated on-chip thermal sensors into the model, which allowed us to generate these images. Figure 4d illustrates the assumed packaging configuration.) Figure 4c shows the localized heating behavior in a multicore chip model (we assume uniform power density within each core).

As Figures 5a and 5b show, the thermal gradients and the peak temperatures in the thinned silicon layers are more pronounced compared to the thicker silicon. This is because of more horizontal heat flow hence, there is a smaller thermal gradient within the thicker layers. The peak temperature is about 50°C higher for the thinned layers at the same power density at the hot spot. Severe heat and large thermal gradients can cause major problems in a processor system. Careful thermal analysis and planning are essential for the viability of a robust 3D stack.

Most of the heat flow within a 3D stack conducts within the silicon. BEOL insulating layers can be excellent thermal insulators as well, so much so that they can completely obstruct the heat flow in some cases. Figure 6 shows a sample cross-section of a stack where we can observe the vertical heat dissipation both within the stack and beyond for a two-layer stack. The temperature difference in this stack is only a few degrees in the vertical direction, but we can see the difference in the layers. The layer closer to the heat sink (top layer) is cooler than the layer below, although the power densities are comparable. The cycles (or it will crack the chip outright if sufficiently heated) because of its large TCE mismatch with silicon. Similar reliability problems can happen with tungsten vias as well, although the mechanisms may be different. There are other metallurgical issues beyond TCE that need resolution before we can definitively conclude what to use.

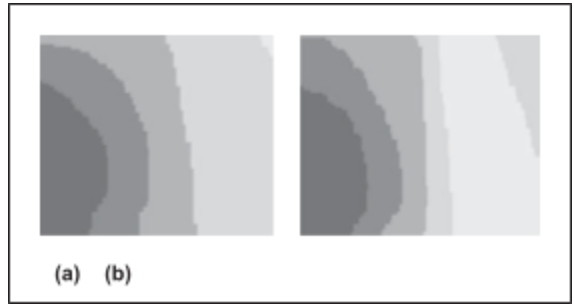

Figure 5. Impact of silicon thickness on thermal gradients. Top view of a simulated chip with a hot spot on the left, using thick silicon (a) and thinned silicon that has half the thickness (b).

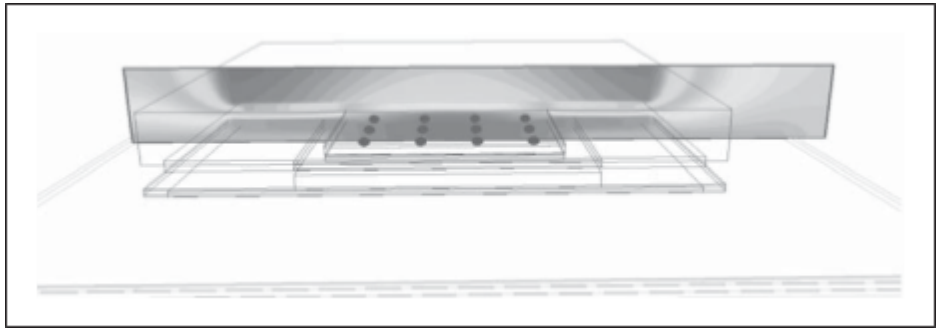

Figure 6. Vertical dissipation of heat across the layers in the stack.

The locations of hot spots in the 3D stack can greatly influence the worst-case temperature and thermal gradients, because the localized hot spots on one layer can exacerbate those on adjacent layers. Thermal planning and management——and a tool infrastructure to support them——are essential for designing high-power systems.

### Physical properties of TSVs

A number of factors impact the characteristics of the vias——including their size and geometry, and the manufacturing process details as well as material selection——but we consider only the first-order material implications here. This is not the total story. In Table 1, we see that copper has much less electrical resistance (about 3×) and better thermal conductivity (more than 2×) than tungsten. Therefore, copper would be the obvious choice for via material, except that the thermal coefficient of expansion (TCE) for copper is about 4× greater than that of tungsten, and more than 6× that of silicon.

Ironically, it's exactly those high-power applications where copper would help the most that cause copper to be the bigger concern from a reliability point of view. That is, while having the least electrical loss and best thermal removal, copper will cause fissures in the silicon over a large number of thermal is in what order to stack the layers. Specifically, if we have an extremely hot layer with high-speed signals (such as processors), and a relatively cool layer with relatively slower signals (such as DRAM), which chip should be closest to the package, and which should be closest to the heat sink? While locating the high-speed layer closest to the package results in better signal integrity (because the signals don't need to pass through as many vias), the thermal resistances of the slower layers between the hot layer and the heat sink makes heat removal harder.

Removing heat is paramount; consequently, you should locate the hottest layer against the heat sink. If this causes unacceptable signal loss, the design should fix this in other ways (for example, it either slows the signals down, or uses repeaters or repeater latches). Further, as we discussed, copper vias (which may have about 2.7× the thermal conductivity of silicon) will cause reliability problems if heat is an issue because of the TCE mismatch. Furthermore, although tungsten may be safer to use in high-power applications (particularly if you're using lots of them), tungsten has only about 15% more thermal conductivity than silicon. Therefore, tungsten vias won't help to remove heat much more than silicon. At this time, we believe the potential for thermal vias to help with heat removal is minimal, and the general promise of thermal vias is not as promising as we had once hoped.

Table 1. Electrical and thermal properties of silicon, silicon dioxide, copper, and tungsten.

| Material | Silicon (Si) | Silicon dioxide ($SiO_2$) | Copper (Cu) | Tungsten (W) |
|---|---|---|---|---|
| Electrical resistivity (n$\mathbf{V}$m at 20°C) | Semiconductor | Insulator | 17 | 53 |
| Thermal coefficient of expansion ($10^{-6}$ m/m — K) | 2.6 | Varies | 16.5 | 4.5 |
| Thermal conductivity (W/m — K) at 300 K | 149 | 10 | 401 | 173 |

* In the "Material" column, n is an integer that depends upon the nature of interaction, m stands for meter, K stands for Kelvin and W stands for watts.

IMPENDING DIFFICULTIES in lithographic scaling, combined with many recent successes in making large numbers of TSVs, makes this time ripe for innovation in 3D silicon technology. Future chips might, in fact, be aggregated conjoinings of multiple (thinned) chip layers. Although this affords some new opportunities and new ways of thinking about the logistics of planning and building future products, it also comes with a new set of challenges.

At first glance, 3D integration may appear to simply be a new manifestation in improved packaging density. Although this is true on a simplistic level, on closer examination, we find that 3D integration has many new nuances: it offers some advantages in the logistics of composing systems, and comes with some completely new design questions that will require new design infrastructures.

Although a natural first reaction to 3D technology is that it must afford a huge performance advantage, we find that with the exception of possibly enabling massive internal bandwidth, the main advantages of 3D integration are logistical. For example, 3D systems enable the physical integration of subsystems, possibly in disparate technologies or across technology generations. If done wisely, subsystem integration enables the volumization of, and reuse potential for, subsystem componentry over a range of products and (perhaps) product generations. Subsystem integration also permits a more elegant incorporation of infrastructure (electrical and test) into a product, which may require some standardization, and which certainly requires more forethought than we ever needed for 2D components.

One of the primary issues (at least, in high- power applications) which drives many subsequent design decisions is the placement, pitch, and material composition of vias. On reflection, this is hardly surprising—the vias are what enable 3D technology in the first place. As we discussed, some new thermal challenges will arise regarding 3D stacks, and we'll need new thermal modeling tools to ensure good thermal design.

Overall, while 3D integration certainly opens up new opportunities and business models for the electronics industry, it comes with a completely new set of challenges—hence, opportunities within the design and test communities. We hope that we've provided insight into what some of those new opportunities and challenges might be.